\documentclass[aps,pre,graphicx,reprint,amssymb,showpacs,superscriptaddress,floatfix,nobalancelastpage]{revtex4-1}
\usepackage[utf8]{inputenc}
\usepackage[T1]{fontenc}
\usepackage{lmodern}					
\usepackage{amsmath}
\usepackage{amsfonts}
\usepackage{amssymb,textcomp}
\usepackage[xcdraw]{xcolor}
\usepackage{colortbl}
\usepackage{tikz}
\usepackage{bm}
\usepackage{framed}
\usepackage{graphicx, wrapfig}
\usepackage{array,multirow}
\usepackage{natbib}
\usepackage[margin=2cm]{geometry}

\usepackage{hyperref}

\newcommand\beq{\begin{equation}}      
\newcommand\beqnn{\begin{eqnarray*}}   
\newcommand\beqa{\begin{eqnarray}}     
\newcommand\beqann{\begin{eqnarray*}}  

\newcommand\eeq{\end{equation}}        
\newcommand\eeqnn{\end{eqnarray*}}     
\newcommand\eeqa{\end{eqnarray}}       
\newcommand\eeqann{\end{eqnarray*}}    

\newcommand{\ave}[1]{\left\langle #1 \right\rangle}

\def\nl {\nonumber \\}

\newcommand\bi{\begin{itemize}}
\newcommand\ei{\end{itemize}}

\def\nl {\nonumber \\}

\def\bx{\bm x}

\newcommand{\eref}[1]{(\ref{#1})}

\newcommand{\fref}[1]{Figure~\ref{#1}}

\newcommand{\al}[1]{\begin{align} #1 \end{align}}
\def\l0{L}

\newcommand{\grad}{\nabla}
\renewcommand{\div}{\nabla\cdot}
\renewcommand{\vec}[1]{\bm{#1}}
\newcommand{\Gvec}[1]{\boldsymbol{#1}}

\newcommand*\colvec[3][]{
    \begin{pmatrix}\ifx\relax#1\relax\else#1\\\fi#2\\#3\end{pmatrix}
}

\begin{document}

\title{Non-equilibrium forces following quenches in active and thermal matter}
\author{Christian M. Rohwer}
\email[]{crohwer@is.mpg.de}
 \affiliation{Max Planck Institute for Intelligent Systems, Heisenbergstr. 3, 70569 Stuttgart, Germany}
 \affiliation{4th Institute for Theoretical Physics, Universit\"at Stuttgart, Pfaffenwaldring 57, 70569 Stuttgart, Germany}
 \author{Alexandre Solon}
 \affiliation{Department of Physics, Massachusetts Institute of Technology, Cambridge, Massachusetts 02139, USA}
 \author{Mehran Kardar}
 \affiliation{Department of Physics, Massachusetts Institute of Technology, Cambridge, Massachusetts 02139, USA}
 \author{Matthias Kr\"uger}
 \affiliation{Max Planck Institute for Intelligent Systems, Heisenbergstr. 3, 70569 Stuttgart, Germany}
 \affiliation{4th Institute for Theoretical Physics, Universit\"at Stuttgart, Pfaffenwaldring 57, 70569 Stuttgart, Germany}

\date{\today}

\begin{abstract}
Non-equilibrium systems with conserved quantities like density or momentum are known to exhibit long-ranged correlations. This, in turn, leads to long-ranged fluctuation-induced (Casimir) forces, predicted to arise in a variety of non-equilibrium settings. Here, we study such forces, which arise transiently between parallel plates or compact inclusions in a gas of particles, following a change (``quench'') in temperature or activity of the medium. Analytical calculations, as well as numerical simulations of passive or active  Brownian particles, indicate two distinct forces:
{\bf (i)} The immediate effect of the quench is adsorption or desorption of particles of the medium to the immersed objects, 
which in turn initiates a front of relaxing (mean) density. This leads to time-dependent {\it density-induced forces}. 
{\bf (ii)} A long-term effect of the quench is that density fluctuations are modified, manifested as transient (long-ranged) (pair-)correlations
that relax diffusively to their (short-ranged) steady-state limit. As a result, transient {\it fluctuation-induced forces} emerge.
We discuss the properties of fluctuation-induced and density-induced forces as regards universality, relaxation as a function of time, and scaling with distance between objects. 
Their distinct signatures allow us to distinguish the two types of forces in simulation data. Our simulations also show that a quench of the effective temperature of an active medium gives rise to qualitatively similar effects to a temperature quench in a passive medium. Based on this insight, we propose several scenarios for the experimental observation of the forces described here. 

\end{abstract}

\maketitle

\section{Introduction}

Inclusions introduced into a fluctuating medium disturb the fluctuations and in turn experience  fluctuation-induced forces (FIFs)~\cite{kardargolestanian1999}. 
A well-known example is the Casimir force between parallel mirrors that constrain {\it quantum fluctuations} of the electromagnetic
field~\cite{casimir1948,bordag}, as well as the related London~\cite{london}, Casimir-Polder~\cite{polder} and van der Waals~\cite{parse} interactions between polarizable objects.
Constrained {\it thermal fluctuations} in a solution of polymers or colloids lead to so-called {\it depletion forces}~\cite{deplete}.
Unlike the Casimir/van der Waals interactions, depletion forces are short-ranged and non-universal, and depend on microscopic
properties of the medium and inclusions. However, as pointed out by Fisher and de Gennes~\cite{fisherdegennes}, thermal FIFs 
become long-ranged and universal due to long-ranged correlations emerging near a critical point~\cite{hertlein2008,gambassiCCFPRE2009,garciachan2002,garciachan2006,fukutowettingfilms2005,linzandi2011}.

While in a typical fluid in thermal equilibrium, long-ranged correlations (and thus long-ranged fluctuation forces) occur only in special circumstances, 
e.g., at the critical point, such correlations are more common out of equilibrium~\cite{grinsteinleesachdev1990}. Indeed, in the presence of conserved quantities (such as density), systems out of equilibrium generically display long-ranged  correlations~\cite{spohn1983,dorfmankirkpatricksengers1994,mukamelkafri1998}. Associated FIFs have been studied theoretically in driven steady-states such as fluids subject to temperature gradients~\cite{kirkpatricksengers2013,kirkpatrick2015prl,kirkpatrick2016pre}, particles diffusing in a density gradient~\cite{aminovkardarkafri2015}, and in shaken granular systems~\cite{wadasasa2003,sotogranular2006,shaebaniwolf2012}. Rather than driven states, Ref.~\cite{rohwer2017transient} considered forces in transient non-equilibrium states following temperature- or activity-quenches. These forces result from \textit{conserved} density fluctuations (``model B'' dynamics~\cite{hohenberg}), and occur when transient long-ranged correlations emerge after a rapid change in temperature or noise-strength. 
Here, we expand on and generalize such transient FIFs.

We note that non-equilibrium FIFs have been discussed in many other
contexts. The prototypical example is radiation pressure due to a flux
of photons, and the associated near-field forces between objects
maintained at different temperatures~\cite{Antezza08,krueger2011}. On
the classical side, various non-equilibrium aspects of critical
Casimir forces have been investigated. These include the force
response to external perturbations~\cite{gambassi2006} or temperature
quenches~\cite{gambassi2008EPJB,deangopinathan2009JStatMech,deangopinathan2010PRE},
vibrating surfaces~\cite{hanke}, moving
objects~\cite{gambassi2013prl}, and for
shear-perturbation~\cite{rohwer2017viscosity}. Non-equilibrium thermal
Casimir forces have also been studied for Brownian
charges~\cite{deanpodgornik2014,lu2015epl,dean2016prl}. In contrast to
the above, we focus on setups where long-ranged forces are absent in
the underlying steady states.

To simplify analytical and numerical studies we focus on systems with
only one conserved quantity, namely the particle number. A well-studied model system is that of passive Brownian particles, which will be underlying most of our theoretical approaches. Another, 
particularly timely example is that of {\it dry active
  matter}~\cite{activeRMP}. Asymmetric patterning of activity of
colloidal particles can lead to self-propulsion~\cite{golestanian},
with collections of such particles exhibiting myriad active phases
which have been subject to intense theoretical~\cite{freyX} and
experimental~\cite{bizonne} investigations. We focus here on the
dilute (gas-like) phase with no emergent symmetry breaking, where
density fluctuations are short-ranged in the steady
state. Nonetheless, the absence of time-reversal
symmetry~\cite{catesX} makes these systems different from a
conventional gas; e.g., they may or may not posses an equation of
state governing the pressure exerted on a boundary. The question of
Casimir-like FIFs for parallel plates inserted in an active gas has
also been explored~\cite{reichhardt2014}. Such forces exist, but, like
depletion forces, are short-ranged, arising from accumulation of
active particles at surfaces.
In these models, the particles undergo
stochastic motion, due to thermal motion or self-propulsion, with
density as the only conserved quantity; momentum and energy are
dissipated to the bath. 

We demonstrate that temperature quenches lead to two
types of forces between objects embedded in a fluid of Brownian
particles: Density-induced forces (DIFs) as well as the
fluctuation-induced forces predicted in
Ref.~\cite{rohwer2017transient}. The two effects have different
origins. We show that DIFs appear because of changes in the {\it mean}
density after the quench, especially in the boundary layer near the
embedded objects because of adsorption and desorption. The (diffusing) change in
density in turn induces a change in the force exerted by the bath, 
leading to a long-ranged interaction between two objects. This type of
interaction has precedent in other non-equilibrium situations: Driving
external objects (such as spheres) through a suspension of Brownian
particles can result in a change in the mean density (e.g. accumulation between the spheres), and thus lead to forces
between the driven
objects~\cite{loewen2003prl,kruegerrauscher2007}. On the other hand,
FIFs are due to non-equilibrium fluctuations and correlations following the quench, and thus appear even if the mean density remains
constant. FIFs appear because of changes in the pair
correlations in the medium and hence rely on interactions between the
particles~\cite{rohwer2017transient}, while, as we will show, DIFs
already appear in the dilute limit.

We investigate the properties of these forces both analytically and in numerical simulations of the above-mentioned model systems after a quench in the temperature or in the activity of the bath particles. Regarding the geometrical setup, we study two typical paradigms: Two parallel plates exemplify the case of closed systems (or non-compact objects), while the case of open systems is investigated via the example of two small (compact) inclusions. 

Summarizing, we find that DIFs and FIFs are superimposed, but can be distinguished due to their different characteristic
signatures. Indeed, while both are long-ranged
(algebraicailly decaying) in space, FIFs are also long-lived (i.e., algebraically decaying in time) whereas DIFs are exponentially
cut off in time. The former effect thus dominates at long times after the
quench while the latter is found to dominate at earlier
times. Although active particles are less amenable to analytical
treatment, DIFs and FIFs appear to arise similarly for active and
passive particles. This opens many possible experimental realizations
in systems where a quench in activity can be implemented. 

In Sec.~\ref{sec:simu}, we introduce the model systems of  passive and active Brownian particles. Starting with  non-interacting particles, we investigate DIFs in Sec.~\ref{sec:ig}. Adding interactions between the fluid particles allow us to study FIFs in Sec.~\ref{sec:FIF}. We close with a summary in Sec.~\ref{sec:con}.


\section{System and simulation details}
\label{sec:simu}

Consider a bath of $N$ overdamped active or passive Brownian particles, so that the dynamics of the $i$-th particle follows the Langevin equation~\cite{fily_athermal_2012}
\begin{align}
\partial_t {\vec x}_i &= v_0\vec u(\theta_i) - \mu_0\sum_{j } \grad_i U(\vec x_i-\vec x_j)-\mu_0 \grad_i V \nl
&\qquad +\sqrt{2\mu_0 k_B T}\Gvec\eta_i, \nl
\partial_t \theta_i &= \sqrt{2D_r}\xi_i,
\label{eq:langevin}
\end{align}
where ${\vec x}_i$ and
${\vec u}(\theta_i)=(\cos\theta_i,\sin\theta_i)^T$ (in 2d) are position and orientation vectors, respectively. The particles interact via a pair potential $U$, while $V$ is the external potential which models the immersed objects (i.e., parallel plates or inclusions; see below). $\mu_0$ is a mobility coefficient, $k_BT$ the thermal energy, $v_0$ the self-propulsion velocity and $D_r$ is the rotational diffusion coefficient. 
$\Gvec\eta_i$ and $\xi_i$ are Gaussian white noises with correlations
\begin{align}
\langle \eta_{i\alpha}(t)\eta_{j\beta}(t')\rangle &= \delta_{ij}\delta_{\alpha\beta}\delta(t-t'), \nl
\langle \xi_{i}(t)\xi_{j}(t')\rangle &= \delta_{ij}\delta(t-t'),
\end{align}
where Roman and Greek letters denote particle indices and Cartesian coordinates, respectively. Numerically, Eqs.~(\ref{eq:langevin}) are integrated using a forward Euler scheme. 

We will consider Eq.~(\ref{eq:langevin}) in two limits: Passive
Brownian particles (PBPs) with $v_0=0$
and $T\neq
0$, and active Brownian particles (ABPs) with $v_0\neq 0$ and
$T=0$.
The two cases are made comparable by introducing the effective
temperature $k_B
T_{\rm eff}=v^2/(2\mu_0
D_r)$ for our ABPs. Indeed, the (large-scale) diffusion coefficient of
a freely-diffusing ABP equals $\mu_0
k_B T_{\rm
  eff}$~\cite{solon_active_2015}. Also, a suspension of ideal ABPs
exerts a pressure $P=\rho_0
k_B T_{\rm
  eff}$ on a planar wall, as in the ideal gas law, irrespective of the
wall potential~\cite{mallory_anomalous_2014,solon_pressure_2015}, were
$\rho_0$ is the density far from the surface.

A temperature quench is implemented by instantaneously changing the
value of $T$ or $v_0$ in Eq.~\eqref{eq:langevin}. The time-independent
states before and long after the quench are equilibrium (PBP) or
steady states (ABP).  For ease of notation and readability of the
paper, in the following we partly omit the subscript `${\rm eff}$' for
$T_{\rm eff}$, as well as the distinction between equilibrium and
steady states.

We consider forces between planar surfaces as well as finite-sized
inclusions. In the simulations, planar surfaces will be modeled by a
repulsive harmonic potential. For example, for the case of a plate at
$z=0$ that confines the fluid to the positive-$z$ side,
\begin{align}
\label{eq:V}
V(z)=\left\{
\begin{array}{cc}
\frac{\lambda_W}{2} z^2,&z<0\\
0,&z>0.
\end{array}
\right.
\end{align}
Inclusions are modeled by a Gaussian potential, see Eq.~\eqref{eq:Vgauss1D} below. The forces acting on the objects (DIFs and FIFs) are unambiguously found by equating the reaction forces on the potential $V$ with the forces exerted on the particles. Naturally, for simulating the ideal gas of BPs in Sec.~\ref{sec:ig}, we set $U=0$.  For the interacting particles simulated in Sec.~\ref{sec:FIF}, we use a short-ranged repulsive potential
\begin{align}
U(r)= \left\{
\begin{array}{cc}
\frac{\lambda}{2} (r-r_0)^2, & r<r_0\\
0,& \mbox{else}.
\end{array}
\right.
\label{eq:U}
\end{align}
Throughout the paper, we avoid crystallization or motility-induced
phase separation~\cite{cates_motility-induced_2015} by considering
small enough $\lambda$ and appropriate ranges of temperatures. Our
systems thus always relax to a homogeneous fluid in steady state.
Simulation units are fixed by setting
$\mu_0=k_B=r_0=1$. Table~\ref{tab:symbols} summarizes important
observables which are considered in the course of the paper.  
Simulations are mostly performed in two spatial dimensions, except for the
one-dimensional simulations of
Figs.~\ref{fig:force1d-obstacles_distance}
and~\ref{fig:force1d-obstacles_convergence}.

As regards theory, quenches of non-interacting media will be studied via the Smoluchowski equation, modelling diffusion of a density of ideal particles in the presence of external potentials \cite{dhont,kreuzer}. Density fluctuations, in turn, are considered in a field-theoretical framework, which arises upon coarse-graining microscopic descriptions \cite{krugerdean2017a}. Theoretical results are presented for spatial dimensions $d=1,$ 2 and 3.

\begin{table}[h]
\begin{tabular}{ |p{2.67cm}||p{5.5cm}|}
 \hline  \hline
 \textbf{Symbol} & \textbf{Meaning}\\
 \hline \hline
 $\hat\rho(\bm x)$ & Density operator $\hat \rho(\bm x,t) = \sum_{i=1}^N \delta(\bm x - \bm x_i(t))$   \\ \hline
 $\rho(\bm x,t)=\ave{\hat\rho(\bm x,t)}$ &  Mean density (in or out of equilibrium)    \\ \hline
 $\Delta\rho(\bm x)$ &  Density adsorbed (or desorbed) at a surface after the quench \\ \hline
 $\rho_0$ &  Density far from surfaces    \\ \hline
 $\phi(\bm x,t)$ &  Fluctuations of density operator about its mean, $\phi(\bm x,t) = \hat\rho(\bm x,t) - \rho(\bm x)$ \\ \hline
 $P(t)$ (plates) &  Pressure acting on the inside surface \\ \hline
 $F(t)$   &  Net force, taking into account the pressures on both surfaces of a plate. Positive force indicates repulsion.\\ \hline
 \hline
\end{tabular}
\caption{Summary of quantities considered in the text.}
\label{tab:symbols}
\end{table}

\section{Quenching an ideal gas: density changes and the associated forces}
\label{sec:ig}

In this section, we consider ideal gases of active or passive Brownian particles, i.e., we set $U\equiv 0$ in Eq.~\eqref{eq:langevin}. 
In the absence of interactions, (pair-)correlations and fluctuations are unaffected by the quench, and the resulting post-quench forces (PQFs) are solely due to changes in mean particle density (i.e., the DIF as denoted above). This statement will be reiterated in Sec.~\ref{sec:FIF} below. It is thus instructive to consider the ideal case first.

Starting from a steady state at a given temperature $T$, the quench initially only affects the {\it boundary layer} near an object like a plate or an
inclusion. Indeed, in equilibrium, the density profile is given by the Boltzmann distribution $\rho(\vec x)\propto \exp(-\beta V(\vec x))$, which depends on temperature via $\beta=1/k_B T$. The fraction of particles adsorbed at the boundary (inside the potential) changes accordingly during a quench, and, due to particle conservation, diffusive fronts are initiated. Pressures and forces are thus time-dependent. For active particles, the same effect is expected, since they form a boundary layer at a surface that depends on the activity of the particles, albeit in a more complicated
manner~\cite{elgetigompper2013active,marchetti2014active,solon_pressure_2015}. Non-interacting active particles show diffusive motion, quantified by the effective temperature $T_{\textrm{eff}}$ introduced in Sec.~\ref{sec:simu} above. In the region close to the surfaces of objects, the ``run length'' of ABPs may give rise to additional phenomena.


In the following, we consider the specific cases of parallel surfaces (Sec.~\ref{sec:twow}) and  inclusions (Sec.~\ref{sec:two-inclusions}). In both cases we provide a coarse-grained analytical  description,  and a comparison to numerical simulations.    

\subsection{Two parallel plates}
\label{sec:twow}

\begin{figure}[t]
\centering
\includegraphics[width=0.9\columnwidth]{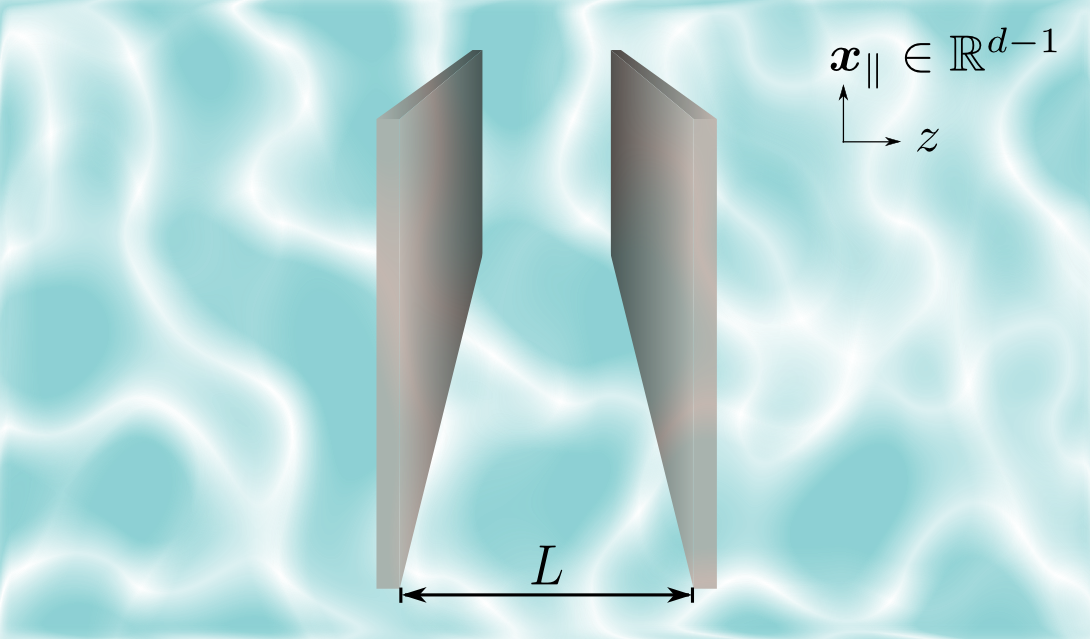}
\caption{Sketch of two parallel plates, separated by a distance $L$
  along the $z$-axis, immersed in a medium. The lateral extensions of the plates are assumed to be much larger than $L$. In Sec.~\ref{sec:twow}, the medium is an ideal gas of active or passive Brownian particles, subject to a quench in temperature or activity. In Sec.~\ref{sec:FIF} the same system with interactions is studied.}
\label{fig:platesSketch}
\end{figure}

We start with the prototypical setup of two parallel plates, separated by a distance $L$ along the $z$-axis; see Fig.~\ref{fig:platesSketch}. $L$ is assumed to be much larger than the width of the boundary layer near the wall so that, in a coarse-grained view, the walls can be modeled as being hard.

\subsubsection{Time-evolution of the density}
Before the quench, the system is assumed to be in a homogeneous state at initial temperature $T_I$ and density $\rho_0$, so that the pressure on the plates is given by the ideal gas law $P=k_BT_I\rho_0$. At time $t=0$, the temperature is switched instantaneously to $T_F$. As argued above, this modifies the boundary layer near the plates, creating an excess or deficit of particles. We thus decompose the mean density $\rho$ between the two plates as
\begin{equation}
  \rho(z,t) = \rho_0 + \Delta\rho(z, t).
\label{eq:rhosplit}
\end{equation}
Note that $\rho$ depends only on $z$ due to translational invariance along $x$ and $y$. In the coarse-grained description, the initial shape of the excess densities is taken as sharp delta-function peaks which model the amount of particles adsorbed or desorbed at the walls,
\begin{align}
  \label{eq:ri}
  \Delta\rho(z,t=0) = \rho_0\left[\alpha_1\delta(z) + \alpha_2\delta(z-L)\right].
\end{align}
Here the adsorption coefficients $\alpha_i$ have units of length, and can be thought of as the change of the width of the boundary layer induced by the quench (see Appendix~\ref{sec:alpha}). For purely repulsive potentials, $\alpha_i<0$ for $T_F>T_I$, and vice versa. If the two surfaces are identical (in terms of their potential), one has $\alpha_1=\alpha_2$. 

For an ideal gas, the excess density $\Delta\rho$ evolves according to the diffusion equation
\begin{align}
  \label{eq:diffusion-deltarho}
\partial_t \Delta\rho(z,t) =D_0 \partial_z^2 \Delta\rho(z,t),  
\end{align}
with diffusion coefficient $D_0=\mu_0 k_B T_F$. The hard walls give rise to no-flux boundary conditions,
\begin{equation}
  \label{eq:no-flux}
 \partial_z \Delta\rho(z,t)|_{z=0} = \partial_z \Delta\rho(z,t)|_{z=L} = 0.
\end{equation}

The solution of Eq.~(\ref{eq:diffusion-deltarho}) for an initial delta-function distribution $g(z,z_0,t=0)=\delta(z-z_0)$, placed at an arbitrary position $z_0$ between the walls, can be written as an infinite sum of image densities placed at $-z_0$, $\pm z_0 \pm 2L$,~$\cdots$, such that 
\begin{align}
  g(z,z_0,t) =
  \sum_{n = -\infty}^\infty \sum_{k = 0}^1G[z,(-1)^kz_0 - 2n L,t]~,
\label{eq:diffsolgen}
\end{align}
in terms of the propagator
\begin{equation}
  \label{eq:diff-propagator}
 G(z,z_0,t) \equiv \frac{1}{\sqrt{4\pi D_0 t}} e^{-(z-z_0)^2/4D_0t}.
\end{equation}
The solution for adsorption/desorption at two surfaces is thus the sum of Eq.~\eref{eq:diffsolgen} with $z_0=0$ and $z_0=L$, so that the excess density is (for later purposes evaluated at $z=0$)
\al{
\Delta\rho(z=0,t) &= \frac {\rho_0} {L} \Big[\alpha_1 \vartheta _ 3\left (0,e^{-\pi^2 t^*} \right) \nl
&\qquad\qquad \quad+  \alpha_2\vartheta _ 3\left (-\pi/2,e^{-\pi^2 t^*} \right)\Big].
\label{eq:sol-deltarho}
}
In the above expression time is rescaled  as  $t^* = D_0 t/L^2$, using the time scale $L^2/D_0$ of diffusion across $L$. $\vartheta_3(u,q) = 1+2\sum_{n=1}^\infty q^{n^2}\cos(2nu)$ is the
Jacobi elliptic function of the third kind~\cite{abramowitzstegun}. The density $\Delta\rho(z=L,t)$ at the second surface is found by interchanging $\alpha_1$ and $\alpha_2$ in Eq.~(\ref{eq:sol-deltarho}).

\subsubsection{Force on the plates}
\begin{figure}[t]
\centering
\includegraphics[width=0.99\columnwidth]{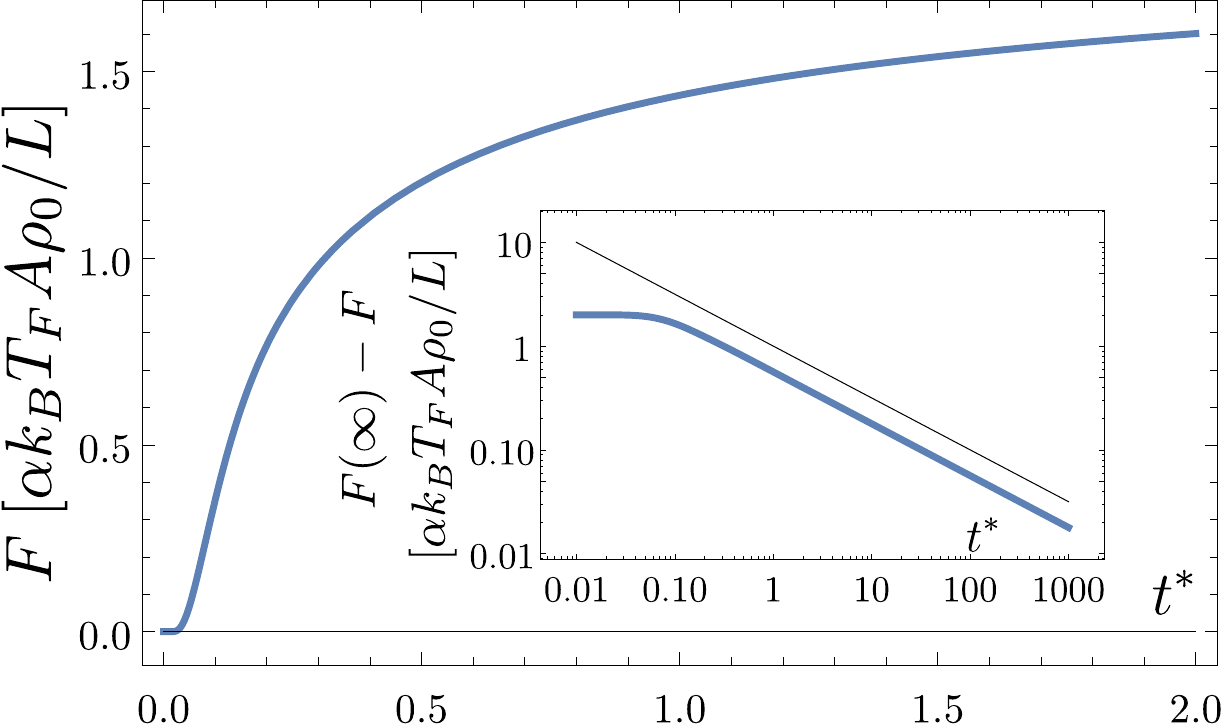}
\caption{Force on the plate at $z=0$ from
  Eq.~(\ref{Fadsorbed}) as a function of dimensionless time
  $t^*= D_0 t/L^2$ for $\alpha_1 = \alpha_2=\alpha_1^{(o)}\equiv\alpha$. The curve approaches 2 for large times. { Inset:} Asymptotic scaling $\sim (t^*)^{-1/2}$ (black
  line) of the difference between the force and its steady state
  value, as given in Eq.~\eref{Fadsorbed2}.}
\label{fig:deltaforce}
\end{figure}

The pressure $P$ exerted on the plate at $z=0$ by the fluid is now directly deduced from Eq.~(\ref{eq:sol-deltarho}), using the ideal gas law 
\al{
  \label{eq:pressure-plate}
P(z=0,t)&=k_BT_F  \left[\rho_0+\Delta\rho(z=0,t)\right]\nl
&=k_BT_F\rho_0+\frac{ k_BT_F\rho_0}{L} \Big[\frac{\alpha_1\vartheta_3\left(0,e^{-1/t^*}\right)}{\sqrt{\pi t^*}} \nl
&\qquad\qquad \quad+ \alpha_2\vartheta_3\left(-\pi/2,e^{-\pi^2 t^*}\right)\Big],
}
The applicability of the ideal gas law in this non-equilibrium situation can be proven starting from Eq.~\eqref{eq:langevin} (for the passive case). Recalling the setup in Fig.~\ref{fig:platesSketch}, we also account for the fluid on the outside surface of the considered plate, by using an adsorption of $\alpha^{(o)}_1$ at its outside face. Assuming a semi-infinite suspension on the outside, we take  $L\to\infty$ in Eq.~\eqref{eq:pressure-plate}, which amounts to using $\lim_{\kappa\to 0}\vartheta_3(0,\kappa)=1$. 
The net force on the plate is then given by the difference of the pressures acting on its two surfaces (a positive force denotes repulsion)
\al{
\frac{F(t)}A
&= \frac{ k_BT_F\rho_0}{L} \Big[ \frac{\alpha_1\vartheta_3\left(0,e^{-1/t^*}\right) - \alpha_1^{(o)}}{\sqrt{\pi t^*}} \nl
&\qquad\qquad \quad+ \alpha_2\vartheta_3\left(-\pi/2,e^{-\pi^2 t^*}\right)\Big],
\label{Fadsorbed}
} where $A$ denotes plate area. (The $\vartheta_3$ function has been
rewritten using Poisson's summation formula; see, e.g.,
Refs.~\cite{apostolbook,borweinbook}.) Equation~\eqref{Fadsorbed} is
independent of the dimensionality of the system, and thus describes
the force between two points in 1d, 2 lines in 2d, or two plates in
3d, as only the variation of density along the $z$ direction is
pertinent. (Note that the bulk contribution $\propto\alpha_1^{(o)}$ is
indeed independent of $L$, as is seen by using $t$ instead of $t^*$.)  As
such, the force scales as $1/L$ in all dimensions, which is in
contrast to the fluctuation-induced force in
Ref.~\cite{rohwer2017transient} (see Sec.~\ref{sec:FIF} below), which
scales as $1/L^d$ in $d$ dimensions. For short times, $t^*\to0$, the
force in Eq.~\eqref{Fadsorbed} vanishes with an essential singularity
if $\alpha_1=\alpha_1^{(o)}$, i.e., if the plate has the same surface
properties on both sides. If $\alpha_1\not=\alpha_1^{(o)}$, $F$
diverges as $1/\sqrt{t^*}$. The singularity is presumably cut off,
depending on details of the potential $V$, which are omitted in this
calculation. For long times, \al{
  \lim_{t^*\to\infty}\frac{F(t)}{A} 
  \notag&=\frac{k_BT_F\rho_0}{L} \Big[\alpha_1 (1+2 e^{-\pi^2t^*}+2
  e^{-4\pi^2t^*}+\dots)\\& -\frac{\alpha_1^{(o)}}{\sqrt{\pi t^*}}\nl
  &+ \alpha_2(1-2 e^{-\pi^2t^*}+2 e^{-4\pi^2t^*}+\cdots)\Big].
\label{Fadsorbed2}
} 
The outside contribution thus relaxes with a power law $\alpha_1^{(o)}/\sqrt{\pi t^*}$ (the semi-infinite space provides no long-time cutoff), while the contribution from between plates relaxes exponentially. We have kept the next-to-leading terms in Eq.~\eqref{Fadsorbed2} in order to demonstrate that for $\alpha_1=\alpha_2$ (i.e., for identical surfaces), this exponential relaxation is particularly fast, since terms describing a potentially slower decay, as  $\sim e^{-\pi^2t^*}$, cancel. Ultimately, the force approaches the steady state value of 
\begin{align}
\frac{F(\infty)}{A}= \frac{k_BT_F\rho_0}{L}(\alpha_1+ \alpha_2),\label{eq:fi}
\end{align}
which resembles the limit were the excess density in Eq.~\eqref{eq:diffusion-deltarho} is distributed homogeneously between the walls.  Note that the contribution $k_BT_F\rho_0$, arising from the bulk density, cancels in the force in Eq.~\eqref{eq:fi}, because it acts on the plate from both sides.

Regarding Fig.~\ref{fig:platesSketch}, there might be another process of diffusion around the edges of the (finite) plates, which may ultimately lead to equilibration of the baths inside and outside the plates. This process is not taken into account here (it is assumed to be much slower than the considered processes).

The force on the plate (Eq.~\eqref{Fadsorbed}) is shown in Fig.~\ref{fig:deltaforce}. The fast initial increase, and the slow power law approach to the final value are clearly discernible. 
\subsubsection{Simulations}
\begin{figure}[t]
\centering
\includegraphics[width=0.99\columnwidth]{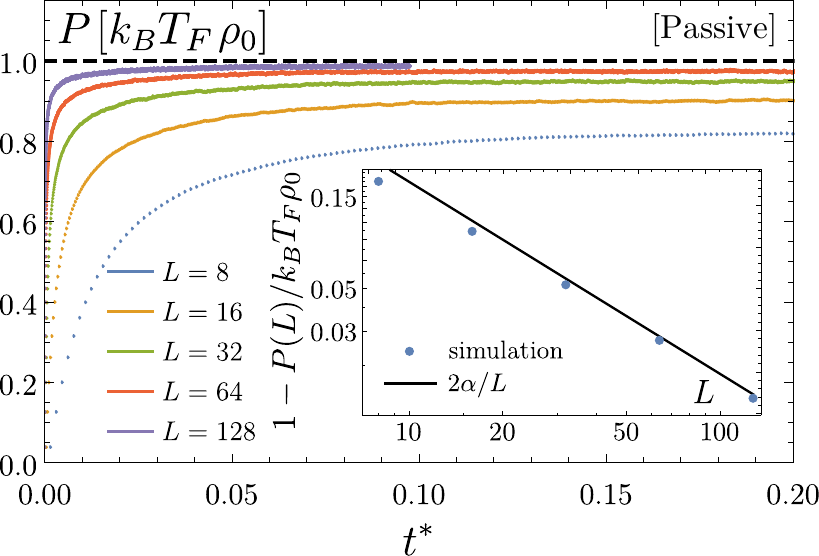}

\vspace{0.3cm}

\includegraphics[width=0.99\columnwidth]{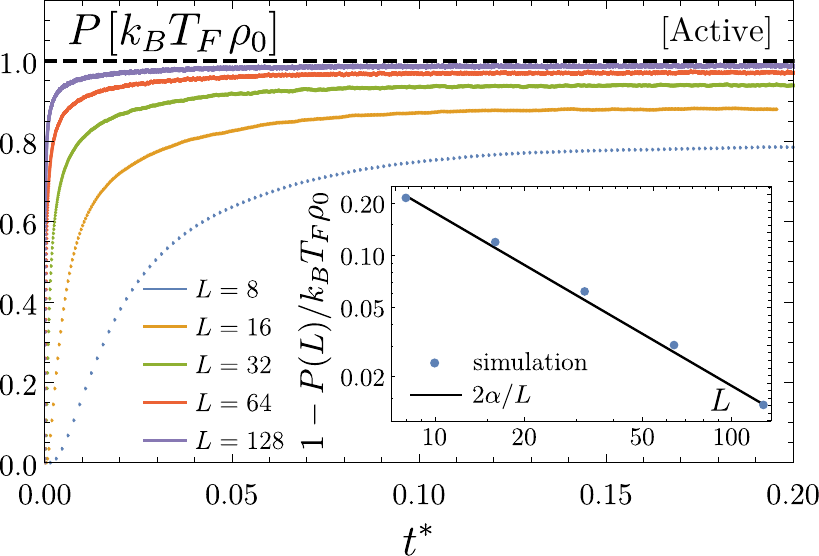}
\caption{Pressure on the plates separated by a distance
  $L$, after a quench from $k_BT_I=0$ to $ k_BT_F=0.5$ (simulation units, see Sec.~\ref{sec:simu}), measured in 2d simulations for PBPs (top) and ABPs (bottom); $t^* = D_0 t/L^2$ with $D_0=\mu_0 k_BT_F$. Due to the adsorbed density, the steady state pressure deviates from $P=k_BT_F\rho_0$ (black dashed line) by a term proportional to $1/L$, see Eq.~\eqref{eq:Pl}. Insets show the $L$-dependence of the final pressure, with the straight lines corresponding to slope $\alpha=-0.89$ found from Eq.~\eqref{eq:calc-alpha}.}
\label{fig:DIF}
\end{figure}

\begin{figure}[t]
\centering
\includegraphics[width=0.99\columnwidth]{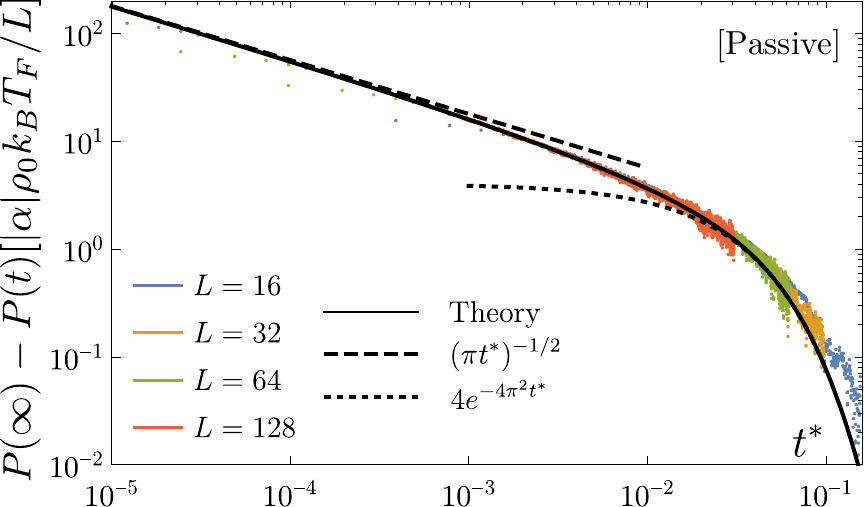}

\vspace{0.3cm}
\includegraphics[width=0.99\columnwidth]{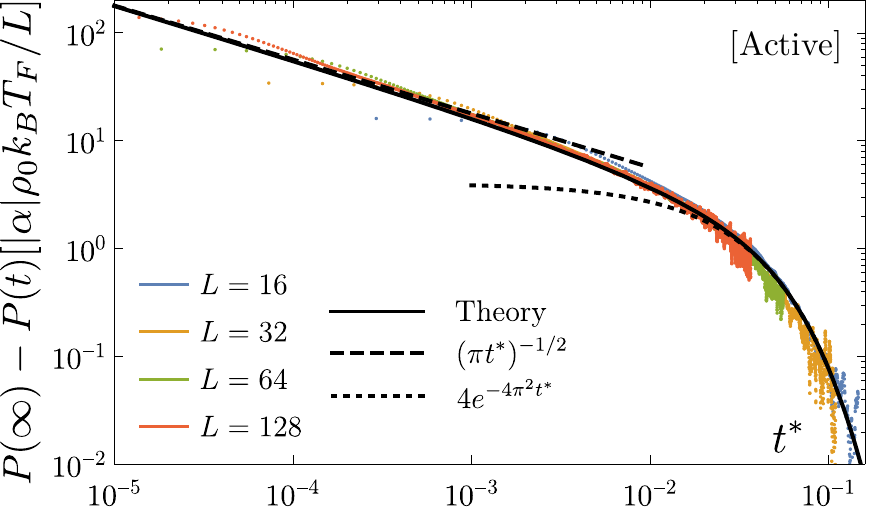}
\caption{Comparison of the data of
  Fig.~\ref{fig:DIF} to Eq.~(\ref{eq:pressure-plate}) (solid black line), for passive (top) and active
  (bottom) particles, with $\alpha=-0.89$ obtained from Eq.~\eqref{eq:calc-alpha}. 
   The dashed and dotted
  lines are the leading behaviors at short and long times, respectively. 
  For the active case, $D_0$ was reduced to $\simeq0.75 \mu_0 k_BT_{\textrm{eff}}$ to obtain agreement.}
\label{fig:DIF_diff}
\end{figure}
Next we compare the above predictions to 2d simulations of both PBPs and ABPs. We consider two identical plates ($\alpha_1 = \alpha_2=\alpha$), and measure the pressure acting on an inside surface, so as to compare to Eq.~\eqref{eq:pressure-plate}. The plates are realized by a quadratic potential (Eq.~\eqref{eq:V} with  $\lambda_W=1$), and we quench from an initial zero (effective) temperature to $T_F$, so that there is no pressure for $t<0$.  The measured pressure after the quench is shown in Fig.~\ref{fig:DIF}. As expected, the steady state pressure reached at long times depends on $L$ in accordance with Eq.~\eqref{eq:pressure-plate}, which in this limit reads  
\begin{align}\label{eq:Pl}
P(t)=k_BT_F\rho_0\left[1+2\frac{\alpha}{L}\left(1+2 e^{-4\pi^2 t^*}+\dots \right)\right].  
\end{align}
The insets of Fig.~\ref{fig:DIF} show the limiting pressure as a function of $L$, which allows a quantitative check of the coefficient $\alpha$. For PBPs, $\alpha$ can be computed explicitly from  the Boltzmann distribution, see Appendix~\ref{sec:alpha}, as
\begin{equation}
  \label{eq:calc-alpha}
  \alpha=\int dz \left( e^{-V(z)/k_B T_I}- e^{-V(z)/k_B T_F}\right),
\end{equation}
where the integration runs over the width of the  surface potential (i.e., where $V_i\neq0$). Equation~(\ref{eq:calc-alpha}) may be interpreted as the difference between the width of the boundary layer in the initial and final states. For the parameters used, this gives $\alpha=-\sqrt{\pi}/2\approx -0.89$, yielding  excellent asymptotes in the insets of Fig.~\ref{fig:DIF}, even for the active case. 

Figure~\ref{fig:DIF_diff} shows the comparison to Eq.~\eqref{eq:pressure-plate} as a function of time. A first confirmation of Eq.~\eqref{eq:pressure-plate} is the very good collapse of data for different $L$ when the pressure is rescaled with $1/L$ and the time axis $t^*$ is used. For short times, the divergence $\sim 1/\sqrt{t^*}$ is observed, from which the simulation data ultimately deviate (due to the short time scale of diffusion across the boundary layer). For long times, the final value is approached exponentially in accordance with Eq.~\eqref{eq:Pl}.


Here, we note a subtlety for ABPs: In order to collapse the data with Eq.~\eqref{eq:pressure-plate} (especially for short times),  
a renormalized diffusion coefficient of $0.75 \mu_0 k_B T_{\textrm{eff}}$ was used. We attribute the necessity to adjust this value to the circumstance that the diffusion coefficient $ \mu_0 k_B T_{\textrm{eff}}$ is only valid in the bulk, and may be expected to be smaller near the walls.

\subsection{Two inclusions at large separations}
\label{sec:two-inclusions}

\begin{figure}[t]
\centering
\includegraphics[width=0.99\columnwidth]{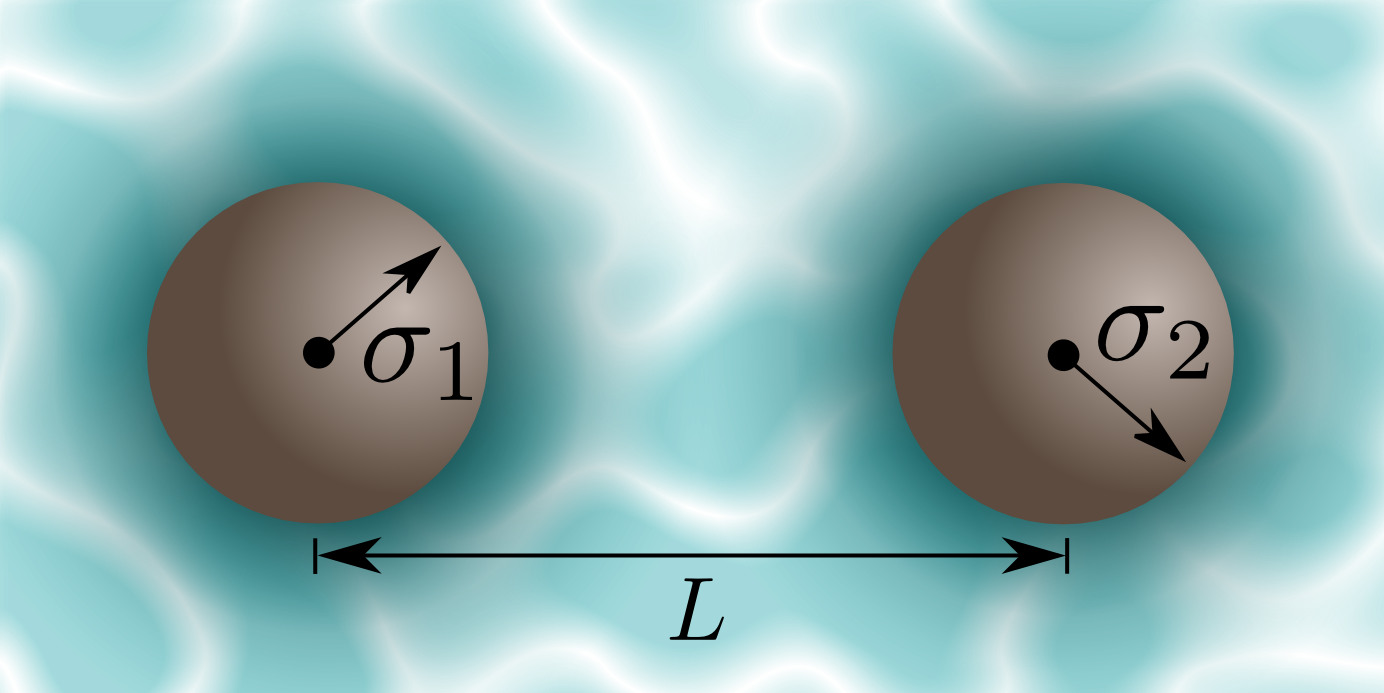}
\caption{Sketch of two inclusions with characteristic sizes $\sigma_i$, separated by a center-to-center distance $L$. The surrounding medium is a suspension of non-interacting active or passive Brownian particles, undergoing a temperature quench. Throughout, the limit $L\gg\sigma_i$ is assumed.}
\label{fig:inclusionsSketch}
\end{figure}

We now study the time-dependent post-quench DIF between two inclusions, modeled by spherically symmetric potentials $V_1$ and $V_2$, immersed in the suspension of ideal PBPs or ABPs at positions $\vec x=0$ and $\vec x=L\vec e_z$, where  $L$ is assumed large compared to the size of inclusions ($\sigma_i$), see Fig.~\ref{fig:inclusionsSketch}. 

As before, we consider a coarse-grained description where the quench leads to a local excess of BPs at the position of the inclusions (mimicking the BPs adsorbed or desorbed by the inclusion-potential). At $t=0$,
\begin{equation}
  \label{eq:deltarho-inclusions}
 \Delta\rho(\vec x,t=0)=\rho_0\left[ \alpha_1\delta(\vec x)+ \alpha_2\delta(\vec x-L\vec e_z)\right]. 
\end{equation}
The parameters $\alpha_i$ are now understood as the change in volume of the boundary layer around the inclusions, and can be computed as (see
Appendix~\ref{sec:alpha})
\begin{align}
\label{eq:alpha2}
\alpha_i=\int d^dx \left(e^{-\beta_I V_i({\bm x}) }-e^{-\beta _F V_i({\bm x}) }\right),
\end{align}
where $V_i$ is the potential of inclusion $i$.

Unlike the plate geometry, the method of images cannot be used to solve exactly for the dynamics with initial condition Eq.~(\ref{eq:deltarho-inclusions}). However, analytical progress is possible in the limit of shallow inclusions, $|V_i|\ll k_B T_{F}$, where $\rho$ is only weakly perturbed by the potentials. The quench creates a disturbance around (say) the first inclusion,
that propagates (in the absence of inclusion 2) as
\begin{align}
  \label{eq:rhoincl}
  \Delta\rho_1(r,t) = \frac{\rho_0\alpha_1}{(4\pi D_0 t)^{d/2}} e^{-r^2/4D_0t},
\end{align}
where $r \equiv |\bm x|$. (We have allowed for an arbitrary dimension $d$). At leading order, the second inclusion experiences the density gradient generated by the first one without influencing it. The force exerted on the second inclusion then reads
\begin{align}
  {\vec F_2}(t) &= \int d^d x  [\nabla V_2({\bm x})] \Delta \rho_1(r,t) \nonumber\\
  \label{eq:Fincl}
  &\approx -\left[\int d^d x   V_2({\bm x})\right] \frac{\partial \Delta \rho_1(r,t)}{\partial r}\Bigr|_{r=L} {\hat {\vec e}}_z,
\end{align}
where ${\hat {\vec e}}_z$ points from the first to the second inclusion, and in the last line we used $\sigma\ll L$ so that $\Delta \rho_1(r,t)$ does not vary on the scale of inclusion 2. Putting Eqs.~(\ref{eq:rhoincl}) and~(\ref{eq:Fincl}) together, we obtain the force (again, positive sign denotes repulsion) 
\begin{equation}
  \label{eq:Fincl-2}
  { F_2}(t) = \frac{\rho_0\alpha_1 \mathcal{V}_2}{2(4\pi)^{d/2} L^{1+d}(t^*)^{1+\frac{d}{2}}}e^{-\frac{1}{4t^*}}
\end{equation}
where we have defined $\mathcal{V}_i=\int d^d x V_i({\bm x})$.

As a complement to the computation yielding Eq.~\eqref{eq:Fincl-2}, the PQF was computed analytically, {\it without explicit coarse-graining}, for inclusions modeled by Gaussian
potentials,
\begin{align}
V_{1,2}(r) = {\frac{V_0 }{(2\pi)^{d/2}}} e^{-\frac{r^2}{2\sigma^2}}\,,
\label{eq:Vgauss1D}
\end{align}
for $k_BT_I\gg V_0$ and $T_F\to\infty$. Equation~(\ref{eq:Fincl-2}) is then recovered in the limit $\sigma \ll L$, see Appendix~\ref{sec:InclAnalyt}.

Returning to generic potentials, we can ask what happens for ``hard'' inclusions, with potentials $|V|\gg k_BT_F$.  For $d\ge 2$, and $\sigma \ll L$, Eq.~\eqref{eq:Fincl-2} still gives the correct  dependence on $L$ and $t$ for hard potentials, as can be argued for by using a multiple reflection expansion (see e.g. Refs.~\cite{dhont,kruegerrauscher2007}).  Equation~(\ref{eq:rhoincl}) is then still expected to hold as the initial density, but the hard inclusion 2 now modifies the density in its vicinity (as corrected for by a reflection term). 
The force remains proportional to the density gradient of $\rho_1$ at the origin of inclusion 2, but the reflection modifies its amplitude, introducing a more complicated dependence on $V_2(\vec x)$. We thus expect a generalization of Eq.~\eqref{eq:Fincl-2} to
\begin{equation}
  \label{eq:Fincl-a}
  { F_2}(t) = a[\beta_FV_2({\bm x})]\frac{\rho_0\alpha \mathcal{V}}{2(4\pi)^{d/2} L^{1+d}(t^*)^{1+\frac{d}{2}}}e^{-\frac{1}{4t^*}}\,,
\end{equation}
involving an amplitude $a$, an unknown functional of the potential, which approaches unity as $\beta_F\to0$.

\begin{figure}[t]
  \centering
  \includegraphics[width=0.99\columnwidth]{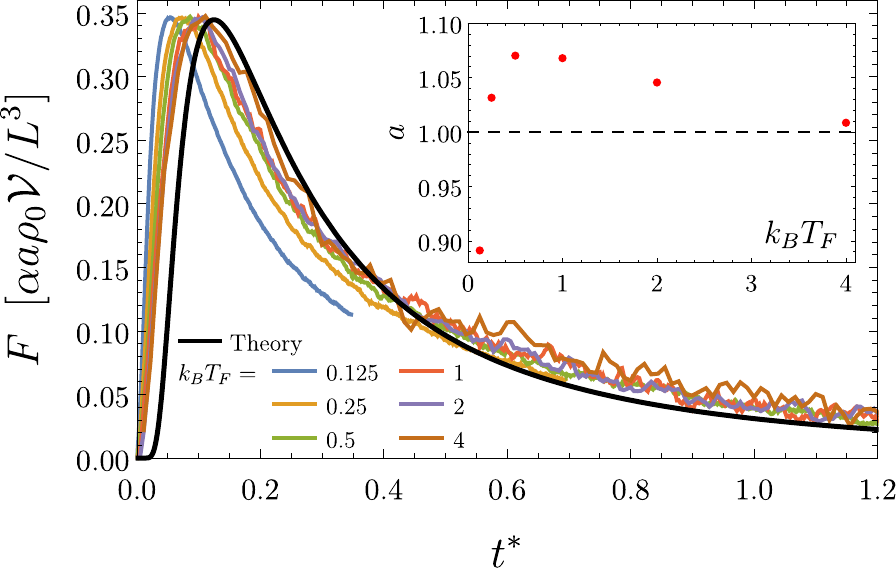}
  \caption{Simulation results of the force between two identical inclusions modeled by the potential in
    Eq.~(\ref{eq:Vgauss1D}) in $d=2$, after a quench from infinite
    temperature to a finite $T_F$ for ideal PBPs. Solid black curve 
    gives Eq.~(\ref{eq:Fincl-a}) using $a=1$. The maxima in the simulation curves have been normalized to unity, which yields the
    temperature-dependent amplitude $a$ shown in the inset.
    $L=6$, $\sigma=1$, $V_0=2\pi$, and system size 
    $72\times 12$ with periodic boundary conditions. The force is repulsive.}
  \label{fig:force2d-obstacles}
\end{figure}

In Fig.~\ref{fig:force2d-obstacles}, we compare simulation results for two inclusions, modeled by the potential in Eq.~(\ref{eq:Vgauss1D}) with $V_0=2\pi$ (simulation units), immersed in an ideal gas of PBPs in 2d. The system is quenched from infinite temperature (a homogeneous initial condition) to a finite temperature $T_F$. $T_F$ is then varied to test Eq.~\eqref{eq:Fincl-a} and to determine $a[\beta_FV_0]$. Equation~(\ref{eq:Fincl-a}) is found to match the simulation data well, except for a shift in time-scale at very low $T_F$. We conjecture this deviation
to be due to corrections of order $\sigma/L$, which become more important at low temperatures. (This effect may be investigated by changing $\sigma$ in Eq.~\eref{eq:Vgauss1D}, which indeed results in a shifted time-scale, as shown analytically in Appendix \ref{sec:InclAnalyt}.) As expected, we find that the amplitude $a$ approaches unity at high $T_F$ (see inset of Fig.~\ref{fig:force2d-obstacles}), so that Eq.~(\ref{eq:Fincl-2}) is recovered in this limit.

\begin{figure}[t]
  \centering
  \includegraphics[width=0.99\columnwidth]{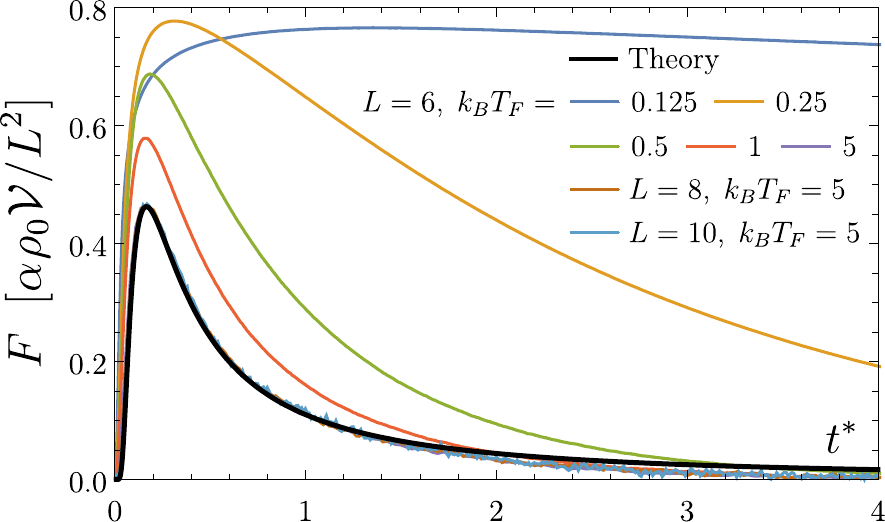}
  \caption{Simulation results of the force between two identical inclusions modeled by the potential in Eq.~(\ref{eq:Vgauss1D}) in $d=1$, after a quench from infinite temperature to a finite $T_F$ for ideal PBPs. The solid black curve  shows Eq.~(\ref{eq:Fincl-2}). At low temperatures, the particles are trapped between the two inclusions which become more and more like impenetrable ``plates'', so that the curve approaches the shape of the curve in Fig.~\ref{fig:deltaforce}.   $\sigma=1$, $V_0=\sqrt{2\pi}$, and system size is $72$ with periodic boundary conditions.}
  \label{fig:force1d-obstacles_distance}
\end{figure}

In $d=1$, the leading orders of reflection do not yield the dominant contribution at large $L$, and the argument leading to Eq.~(\ref{eq:Fincl-a}) does not apply. Figure~\ref{fig:force1d-obstacles_distance} shows that the PQF agrees with Eq.~(\ref{eq:Fincl-2}) at high $T_F$ as expected, with the only visible deviation for long times, were the simulation data approach zero faster than expected from Eq.~(\ref{eq:Fincl-2}). The finite size of the simulation box cuts off the power law decay in time, see Appendix~\ref{sec:convergence-inclusions}. In contrast to the 2d case, the PQF is qualitatively different for low $T_F$, where it tends to the result of Sec.~\ref{sec:twow} for two plates: In this limit particles cannot pass the inclusions (which then, in 1d, become impenetrable ``plates''). Indeed, the summation of image densities performed in Sec.~\ref{sec:twow} can be seen as a reflection expansion, albeit to infinite order.

\section{Fluctuation-induced forces}
\label{sec:FIF}
In the previous section, we noted that temperature or activity quenches lead to excess adsorption or desorption at surfaces, resulting in density ``waves'' and corresponding PQFs, even in the absence of particle interactions. These forces depend on details of potentials characterizing surfaces or inclusions (e.g. via adsorption coefficients $\alpha$). Importantly, in our simulations of ideal gases investigated thus far, we did not see any hint of the post quench fluctuation-induced force predicted in Ref.~\cite{rohwer2017transient}. In this section we demonstrate that such FIFs do occur for \textit{interacting} BPs, via simulations employing a non-zero interaction potential $U$ in Eq.~\eqref{eq:langevin} (see Eq.~\eqref{eq:U}). For concreteness, we focus on the parallel plate geometry of Fig.~\ref{fig:platesSketch}. As we shall demonstrate, a non-zero $U$ is necessary for FIFs to occur, as they are related to 
equilibration of (pair-)correlations of particles. Another insight is that for the system investigated, the DIFs studied in the previous section, and the FIFs considered here, are to a good approximation independent, and are  nearly additive. 

As a reminder, in Sec.~\ref{sec:field-theo}, we expand on the results of Ref.~\cite{rohwer2017transient}, by including quenches between arbitrary initial and final temperatures. In Sec.~\ref{sec:force}, the relation between correlations and the non-equilibrium forces is discussed. 
Finally, in Sec.~\ref{sec:FIF-simus} we identify FIFs in simulation data of both passive and active BPs. 

\subsection{Field theory}
\label{sec:field-theo}

\subsubsection{Preliminaries and static correlations}
We describe density fluctuations in terms of the field $\phi(\bm x,t) = \hat\rho(\bm x,t) - {\rho_0}$, where in this section, we neglect any deviation of $\rho$ from the bulk density $\rho_0$. Coarse-graining beyond any fluid length scales, the resulting Hamiltonian contains only one term~\cite{hohenberg,kardarbook,rohwer2017transient},
\begin{align}
\label{eq:H}
H[\phi] = \int d\bm x\; \frac {m} 2\phi^2(\bm x).
\end{align}
The ``mass'' $m$ in the passive case is given by~\cite{Chandler93,krugerdean2017a},
\begin{equation}
  \label{eq:mass}
  m=k_BT\left(\frac{1}{\rho_0}-c^{(2)}\right)\,,
\end{equation}
where $c^{(2)}$ is the zero wave-vector limit of the so-called direct pair correlation function~\cite{HansenMcDonald}, which is related to the compressibility~\cite{HansenMcDonald} (see Eq.~\eqref{eq:c2} below). In steady state, the Hamiltonian of Eq.~\eqref{eq:H} leads to correlations of density fluctuations,
\begin{equation}
  \label{eq:corre-phi-ss}
\langle \phi(\vec x)\phi(\vec x')\rangle =\frac{k_BT}{m}\delta^d(\vec x-\vec x')\,,
\end{equation}
which are purely local. Consequently, no FIFs are observed in the steady state.

\subsubsection{Post-quench correlations}
Since  the density of particles is conserved locally, the evolution of the field following a quench must be described by a model B~\cite{hohenberg,kardarbook} dynamics. This leads to the stochastic diffusion equation 
\begin{align}
  \label{eq:Lan}
\partial_t \phi(\bm x,t) &=   \div\left(\mu \nabla \left[m \phi(\bm x,t)\right] + \sqrt{2 k_BT\mu}\Gvec \eta \right)\,,
\end{align}
with mobility $\mu$, for which the mapping to Eq.~\eqref{eq:langevin} yields $\mu=\rho\mu_0$ (see e.g. Ref.~\cite{dean1996,krugerdean2017a}). The noise is correlated as
\begin{align}
\langle \eta_{\alpha}(\vec x,t)\eta_{\beta}(\vec
x',t')\rangle=\delta_{\alpha\beta}\delta^d(\vec x-\vec
x')\delta(t-t').
\end{align}

\begin{figure}[t]
  \centering
  \includegraphics[width=0.99\columnwidth]{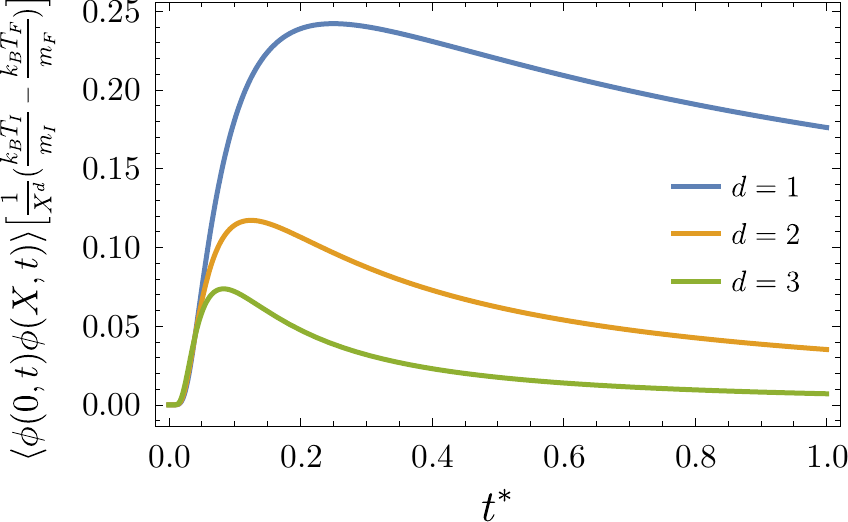}
  \caption{The time-dependent bulk correlation function, relating two points separated by a distance $X$, as given in Eq.~\eref{eq:bulkC(t)}. We show different dimensions $d$.}
  \label{fig:C(t)}
\end{figure}

To compute the post-quench correlation function, we denote $R(\bx,\bx') = -\mu \nabla^2\delta(\bx-\bx')$~\cite{deangopinathan2010PRE,deangopinathan2009JStatMech} and $\Delta_{I/F} = m_{I/F} \delta(\bx-\bx')$, with subscript $I$ and $F$ indicating pre- and post-quench quantities, respectively, as before. (The mass in Eq.~\eqref{eq:mass} is temperature-dependent via the prefactor, but also because $c^{(2)}$ depends on $T$). The time-dependent correlation function for $\phi$ at time $t$ after the quench can then be written as~\cite{deangopinathan2010PRE}
\begin{align}
\langle \phi(\vec x,t)\phi(\vec x',t)\rangle &= k_B T_I e^{-2 t
    \Delta_F R}\Delta_I^{-1}(\bx,\bx') \nl &\quad + k_B T_F
  \Delta_F^{-1}\Big[ 1-e^{-2 t \Delta_F R} \Big](\bx,\bx').
\label{eq:Cxxp2}
\end{align}
We extract from Eq.~\eqref{eq:Cxxp2} the long-ranged parts (which generate long-ranged forces) by noting that $\Delta$ is a local, gradient-free operator, so that only the terms with exponentials yield non-local contributions.  These long-ranged correlations are transient, vanishing as $t\to \infty$.
It is instructive to consider the explicit result for the bulk first, where ($X\not=0$),
\begin{align}
\ave{\phi(0,t)\phi(X,t)}^{LR}
&= \Big[\frac{k_B T_I}{m_I} - \frac{k_B T_F}{m_F}\Big] \frac{1 }{ X^d }\frac{e^{-\frac{1}{8 t^*}}}{(8 \pi  t^*)^{d/2}}.
\label{eq:bulkC(t)}
\end{align}
This equation generalizes the result of Ref.~\cite{rohwer2017transient}, where $T_I=0$ was considered. The time-dependent amplitude of the correlation function is shown in Fig.~\ref{fig:C(t)}.

Equation~\eqref{eq:bulkC(t)} encodes the important result that the temperature quench yields transient long-ranged correlations. Let us recall that these are absent in equilibrium. The physical reason for these correlations is the conservation of particles, which translates to conservation of density. Here,  $t^*= \mu m t/X^2$, i.e., the time scale of diffusion across the distance $X$. These correlations initially rise from zero to a maximal value at around $t^*=1$, but then relax slowly as a power law for large times.

The Casimir forces (FIF) resulting from the correlations in Eq.~\eqref{eq:bulkC(t)} are found by solving Eq.~\eqref{eq:Cxxp2} for two parallel plates with no-flux boundary conditions; see Fig.~\ref{fig:platesSketch}. (Also see, e.g., Refs.~\cite{dietrichdiehl1983,diehljanssen1992,diehlwichmann1995} for details regarding boundary conditions at surfaces in Model B.) The corresponding operator $e^{-2 t \Delta_F R}$ can be computed~\cite{rohwer2017transient}, and yields (with the parallel coordinates set equal, $\vec{x}_\parallel=\vec{x}_\parallel'$, and with $t^*= \mu m t/L^2$)
\begin{align}
  \big\langle\phi(z=0,t)&\phi(z'=0,t)\big\rangle^{LR}  =\notag\\
  &\Big[\frac{k_B T_I}{m_I} - \frac{k_B T_F}{m_F}\Big] \frac{2\vartheta _3\left(0,e^{-\frac{1}{2 t^*}}\right)}{ L^d (8\pi t^*)^{d/2}}.
\label{eq:Cxxp2LR-explicit}
\end{align}
This equation generalizes the result of Ref.~\cite{rohwer2017transient} to include non-zero $T_I$. While this correlation function diverges for $t^*\to0$ (due to coinciding points), 
it relaxes as a power law for large times (with a reduced power compared to Eq.~\eqref{eq:bulkC(t)}) as
\begin{align}
\lim_{t^*\to\infty}\frac{2\vartheta _3\left(0,e^{-\frac{1}{2 t^*}}\right)}{(8\pi t^*)^{d/2}}=  
{(8 \pi  t^*)^{-(d-1)/2}}\, .
\end{align} 
\subsubsection{No non-equilibrium correlations (and thus no FIFs) in an ideal gas}

For an ideal gas (i.e., $U=0$ in Eq.~\eqref{eq:langevin}), the direct correlation function is zero,  $c^{(2)}_{I/F}=0$~\cite{HansenMcDonald} (see also Eq.~\eref{eq:c2} below), so that, from
Eq.~(\ref{eq:mass}) we have $m_{I/F}=k_BT_{I/F}/\rho_0$. The non-equilibrium long ranged correlations in Eqs.~\eqref{eq:bulkC(t)} and \eqref{eq:Cxxp2LR-explicit} are thus zero, and the effect of FIFs (to be discussed below) is consequently absent. This confirms and explains the fact that no FIFs were observed in the simulation data for ideal gases presented in Sec.~\ref{sec:ig}.



\subsection{Fluctuation-induced force}
\label{sec:force}
Equation~\eqref{eq:Cxxp2LR-explicit} is the non-equilibrium transient correlation function for two parallel plates, evaluated at one of the surfaces. A non-trivial step is the computation of local forces or pressures from this correlation function. We present two approaches in this subsection. 

\subsubsection{Force from Gaussian field theory}
For the Gaussian field theory, the stress tensor of the Hamiltonian in Eq.~\eqref{eq:H} is given by ${\boldsymbol \sigma}=-\delta_{ij} \frac{m \phi^2}{2}$~\cite{krugerdean2017b}. 
Using this, the result of Ref.~\cite{rohwer2017transient} for the FIF-contribution to the pressure exerted by the fluid on the wall is obtained as
\begin{equation}
  \label{eq:P-CLR}
  P_{\rm FI}(t)=\frac{m_F}{2} \ave{\phi(z=0,t)\phi(z'=0,t)}^{LR}.
\end{equation}
We consider two explicit cases: {\bf (i)} In simulations we set $T_I\to\infty$ (i.e., $c^{(2)}_I=0$), and measure the pressure exerted on 
the inside of one of the surfaces. In this case
\begin{equation}
  \label{eq:PG}
  P_{\rm FI}(t)=-c^{(2)}_F \rho_0\frac{k_B T_F}{L^d } \frac{\vartheta _3\left(0,e^{-\frac{1}{2 t^*}}\right)}{(8\pi t^*)^{d/2}}.
\end{equation}
$c^{(2)}_F$ is the direct correlation function introduced in Eq.~\eqref{eq:mass} above. 
While this expression diverges as $t^*\to0$, this regime is not  relevant to the simulations, as the pressure is dominated by the DIF for short times.

{\bf (ii)} We repeat the case considered in Ref.~\cite{rohwer2017transient} with $T_I=0$, and compute the net force acting on the plate by taking into account the pressure acting on the outside face due to the bulk ($L\to\infty$) system. Thus,
\al{
\frac{F_{FI}(t)}{A}= \frac{k_B T_F}{L^d} \frac{\vartheta _3\left(0,e^{-\frac{1}{2 t^*}}\right)-1}{ (8\pi t^*)^{d/2}} .
\label{eq:ForceP}
}
In contrast to Eq.~\eqref{eq:PG}, this force approaches zero for $t^*\to0$, since the short-time divergence is a bulk property of the medium which cancels in the subtraction in Eq.~\eqref{eq:ForceP}. Note that the force amplitude of Eq.~\eqref{eq:ForceP}  does not depend on microscopic details such as $\rho_0$. This is because for $T_I=0$, the dependence on $m_F$ cancels; for $T_I\neq0$ and $T_F\neq0$, the force (or pressure) generally depends explicitly on the masses $m_{I/F}$.

\subsubsection{Force from a local equilibrium assumption}
\label{sec:local-equ}
The expression for the force in Eq.~\eqref{eq:P-CLR} relies on the stress tensor from the Gaussian theory. Since realistic systems may display non-Gaussian fluctuations (e.g., the distribution for an ideal gas is Poissonian~\cite{velenich08}), it is useful to consider an alternative method for relating the density correlations in Eq.~(\ref{eq:Cxxp2LR-explicit}) to local pressures. Indeed, the non-equilibrium (long-ranged) modes of $\phi$ decay slowly compared to the fast local ones, so that it is reasonable to assume that the system is locally in equilibrium at the density $\rho_0+\phi(\vec x,t)$. The pressure can then be found from the equilibrium (or steady state for ABPs) equation of state, $P_{ss}(\rho)$, expanded in order to account for fluctuations. At lowest contributing order,
\begin{align}
P[\rho_0+\phi]=P_{ss}({\rho_0}) + \frac{P_{ss}''({\rho_0})}{2}\left[\langle\phi^2\rangle-\langle \phi^2 \rangle_{ss}\right],
\label{eq:P-exp-LEA} 
\end{align}
(See Ref.~\cite{solon_pressure_2015PRL} for the equation of state of an active fluid.) Similar approximations have been proposed in Refs.~\cite{Ortiz,sotogranular2006,kirkpatricksengers2013,aminovkardarkafri2015}. The second term in Eq.~\eqref{eq:P-exp-LEA} subtracts $\langle \phi^2 \rangle_{ss}$  to avoid double counting of (local) steady state fluctuations of Eq.~\eref{eq:corre-phi-ss}. This difference on the right of Eq.~(\ref{eq:P-exp-LEA}) is thus the non-equilibrium part of the correlator considered in the previous subsection, and the fluctuation-induced pressure acting on the wall at $z=0$ follows as
\begin{equation}
  \label{eq:P-FI-LEA}
  P_{\rm FI}(t)=\frac{P_{ss}''({\rho_0})}{2}\ave{\phi(z=0,t)\phi(z'=0,t)}^{LR}.
\end{equation}
Compared to Eq.~(\ref{eq:P-CLR}), the amplitude of the FIF with the local equilibrium assumption is proportional to the second virial coefficient $P_{ss}''({\rho_0})$ instead of the mass $m$, so that the pressure on the inside face after quenching from $T_I=\infty$ is 
\begin{align}
\label{eq:Ple}
  P_{\rm FI}(t)= -\frac{P_{ss}''({\rho_0}) c^{(2)}_F \rho_0}{m_F} \frac{k_B T_F}{L^d } \frac{\vartheta _3\left(0,e^{-\frac{1}{2 t^*}}\right)}{(8\pi t^*)^{d/2}}.
\end{align}

Equations~\eqref{eq:PG} and \eqref{eq:Ple} thus provide alternative results for the same physical situation, which will now be tested in simulations.

\subsection{FIF in simulations}
\label{sec:FIF-simus}
\begin{figure}
  \centering
  \includegraphics[width=0.99\columnwidth]{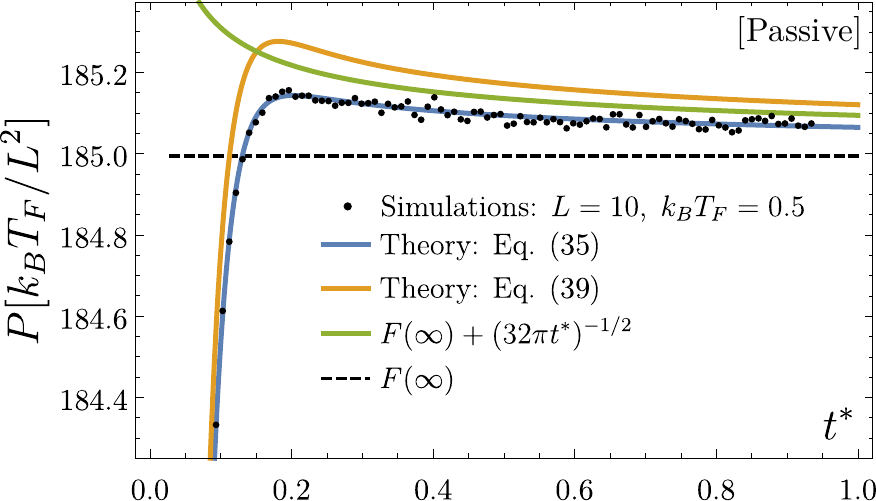} 
  
  \vspace{0.2cm}
  
  \includegraphics[width=0.95\columnwidth]{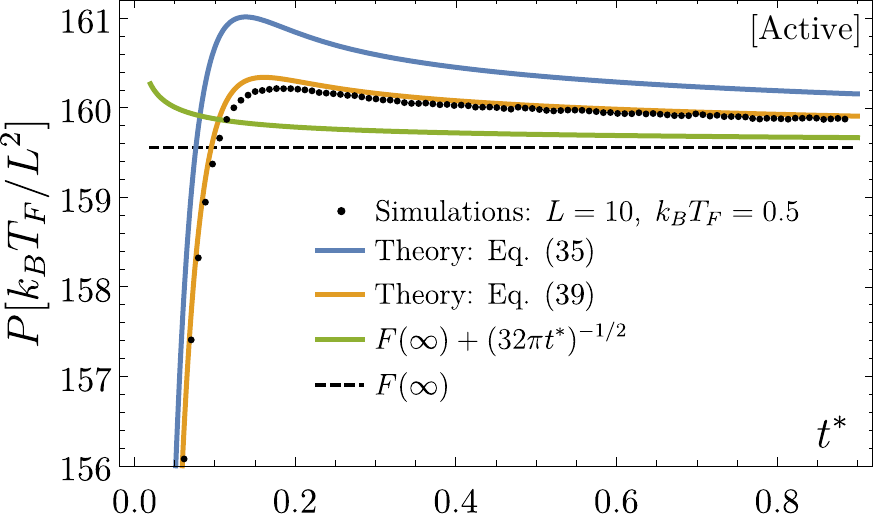}
  \caption{Pressure exerted on the inside surfaces of two parallel plates after a temperature quench to $T_F=0.5$ from simulations (data points). The only difference to Fig.~\ref{fig:DIF} is the presence of an interaction potential $U$. {\bf Top:} PBPs with $\lambda=1$ and $\rho_0=0.936$, giving $P''(\rho_0)=0.674$ and $c^{(2)}=-1.35$. $t^*=\mu m/L^2$ is found with Eq.~\eqref{eq:mass}.
    {\bf Bottom:} ABPs with $\lambda=2$ and $\rho_0=0.909$, giving $P''(\rho_0)=4.06$ and $c^{(2)}=-3.61$. 
    For both: System size $10\times 2000$. The data is averaged over $6\times 10^4$ runs and the final pressure is measured in longer simulations $t^*>200$. The theory curves are the sum of Eq.~\eref{eq:Pl} (with $\alpha$ chosen for best fit) and Eqs.~\eref{eq:PG} or \eref{eq:Ple}, respectively. 
}
  \label{fig:FIF}
\end{figure}

To look for the FIF in simulations, we quench a collection of BPs from infinite temperature (randomly distributed  in the space $0<z<L$) to a finite (effective) temperature of $T_F=0.5$. The resulting pressure acting on an internal surface is shown in Fig.~\ref{fig:FIF}. This graph should be compared to Fig.~\ref{fig:DIF}, where the only difference is the presence of the interaction potential $U$. For short times, the curves in the respective graphs look similar, but there is a pronounced difference for larger times: While the ideal gas curves of Fig.~\ref{fig:DIF} approach their final value from below (and exponentially fast), the curves in Fig.~\ref{fig:FIF} cross zero, and then approach the final value as a slow power law in time. The difference can be identified with the FIF of Eq.~\eqref{eq:PG}, a conclusion which we aim to  make quantitative.

To this end, we first determine the virial coefficients $P_{ss}'(\rho_0)$ and $P_{ss}''(\rho_0)$ independently from simulations in the bulk with periodic boundary conditions as in Ref.~\cite{solon_pressure_2015PRL}. The former yields the direct correlation function~\cite{HansenMcDonald} via (using  $m=(\rho_0^2\chi_T)^{-1}$ with the compressibility $\chi_T=\frac{1}{\rho_0 P_{ss}'(\rho_0)}$)
\begin{equation}
  \label{eq:c2}
  c_F^{(2)}=\frac{1-P_{ss}'(\rho_0)/(k_BT_F)}{\rho_0 }.
\end{equation}
With $c_F^{(2)}$, the mass (which also enters $t^*$) follows via Eq.~\eqref{eq:mass}, so that Eq.~\eqref{eq:PG} can be evaluated \textit{without any free parameters}.  The same is true for Eq.~\eqref{eq:Ple} with additional input of $P_{ss}''(\rho_0)$. 

These equations are shown by lines in Fig.~\ref{fig:FIF}. For the passive case, we see that Eq.~\eqref{eq:Ple} matches simulation data perfectly, while the pressure from the Gaussian stress tensor, Eq.~\eqref{eq:PG}, is slightly off. We have also added the asymptote of Eq.~\eqref{eq:ForceP},  $(32\pi t^*)^{-1/2}$, which also matches quite well, partly by coincidence. However, it  shows that this simple expression yields a good estimate of the FIF-amplitude. The observation that the amplitude of the force in Fig.~\ref{fig:FIF} is of the same order of magnitude as Eq.~\eqref{eq:ForceP} is insightful: The force given by Eq.~\eqref{eq:ForceP} is universal, in the sense that the prefactor depends only on temperature, but not on details of the system. While we discuss corrections to Eq.~\eqref{eq:ForceP} (e.g. due to a finite initial temperature), the shown agreement is still worth noting: FIFs appear to be to some extent universal, in the sense that their order of magnitude can be estimated without knowledge of system details.

At short times, the PQF in Fig.~\ref{fig:FIF} is very similar to
Fig.~\ref{fig:DIF}, as it results from equilibration of the density to
the new temperature. We have thus simply added the predictions for
FIFs to the predicted value for DIFs given by Eq.~\eqref{eq:Pl} with a
fitted coefficient $\alpha$. The resulting blue curve
shows good agreement with simulation data over the whole range shown (for the passive case). The values of $\alpha$ are $-0.41$ (passive) and $-0.62$
(active) which are smaller than in Fig.~\ref{fig:DIF}, because
interactions reduce the amount of adsorbed particles due to
exclusion. Importantly, the observation that Eq.~\eqref{eq:Ple} agrees quantitatively with
the simulation for PBPs demonstrates that the DIFs and FIFs can be
considered quantitatively additive and independent, at least for the
system investigated. However, this is not quite true for ABPs, where,
as in Fig.~\ref{fig:DIF_diff}, the time scale $t^*$ appears difficult
to determine a priori. While a naive estimate of the small wave-vector
diffusion coefficient via $\mu m$ (with Eqs.~\eqref{eq:mass} and
\eqref{eq:c2}) yields $\simeq2.14$ in simulation units, we instead
obtained best agreement by using a value of $1.07$ (roughly half this
value). Furthermore, the Gaussian stress tensor appears in best
agreement with the data, which may be coincidental. To a good
approximation, additivity of the DIFs and FIFs is nonetheless also
displayed for ABPs.

Apart from these issues regarding quantitative description of ABPs, it is worth pointing out that both active and passive BPs show the non-trivial transient fluctuation-induced force, thus confirming the theoretical prediction of Ref.~\cite{rohwer2017transient}. This core finding opens the possibility for experimental detection in many systems, as detailed in Sec.~\ref{sec:con}. While we may expect that DIFs (scaling as $1/L$) in general dominate FIFs (scaling as $1/L^d$), the DIFs in Fig.~\ref{fig:FIF} decay exponentially quickly, so that the power law of FIFs dominates for long times, where it becomes relevant and detectable.

\section{Discussion and Outlook}
\label{sec:con}
\subsection{Conclusions}

\begin{table}[t]
\begin{tabular}{|l||l|p{3.29cm}|}\hline
\hline
\textbf{System}  & \textbf{DIF}& \textbf{FIF}\\
\hline \hline
\multirow{4}{*}{}Parallel  & ${P(t)} \overset{t\to\infty}{\longrightarrow} \rho_0 k_BT_F [\,1 $ & $P(t)\overset{t\to\infty}{\longrightarrow} $ \\
Plates& $+\frac{8\alpha}{L} (1-e^{-4\pi^2t^*})]$ & $ \sim2\frac{k_BT_F}{L^d_{\textrm{\;}}} (8 \pi  t^*)^{-\frac{d-1}{2}}$\\ 
\cline{2-3}
    & Exponential decay& Scale-free algebraic\\
    & in time& decay in time \\ \cline{2-3}
    & Residual pressure &  Vanish at\\
&in steady state&long times\\ 
\hline
\multirow{4}{*}{}Inclusions & $F(t)=$ &~\cite{rohwer2017transient}$F(t)= $ \\
    & $\frac{\alpha\tilde\alpha \rho_0 k_BT_F}{2\sqrt{4\pi}^d}\frac{e^{-\frac{1}{4t^*}}}{L^{1+d}_{\textrm{\;}}(t^*)^{1+\frac{d}{2}}} $ & $\alpha^2\frac{k_B T_F }{L^{2d+1} }e^{-\frac{1}{2 t^*}}\times$ \\
    &  & $\begin{cases}
\frac{(1-t^*)}{16\sqrt{2\pi}\left( t^*\right)^{5/2}},{\;\;\quad\scriptstyle d=1},\hspace{-1cm} \\
\frac{[1-t^*(3t^* + 4)]}{256\sqrt 2 \pi^{5/2}\left( t^*\right)^{9/2}},{\scriptstyle d=3}. \hspace{-1cm}
\end{cases}$
    \\\cline{2-3}
    & No sign change of & Sign change of force \\
    &force & at $t^*\approx 1$ \\\hline\hline
\end{tabular}
\caption{Comparison of density-induced forces (DIFs) and fluctuation-induced forces (FIFs) for objects a distance $L$ apart, where $t^*$ is the correspondingly rescaled time. Expressions are partly simplified for clarity. For parallel plates, $\alpha$ has units of length, and the dimensionless pre-factor for FIFs is of order unity and depends on whether the Gaussian stress tensor or the local equilibrium assumption is employed. For inclusions, $\alpha$ and $\tilde \alpha$ have units of volume (note that their definition differs slightly for FIFs~\cite{rohwer2017transient}). Temperatures may be replaced with effective temperatures for active particles.}
\label{tab:compare}
\end{table}



Rapidly changing the (effective) temperature of active or passive Brownian particles leads to two rather distinct phenomena, which are both due to local density conservation: {\bf (i)} Near immersed objects or boundaries, the temperature quench changes the amount of adsorbed or desorbed particles, so that diffusive fronts are initiated, leading to  density-induced forces (DIFs). For non-interacting particles, this is the only effect which arises after the quench, and it can be described quantitatively using the diffusion equation, for parallel walls as well as for inclusions. For parallel walls, the mean density profile relaxes exponentially quickly between the plates, and the force scales with inverse separation, as $1/L$. The magnitude of the DIF depends explicitly on the potential of the immersed objects.

{\bf (ii)} For interacting BPs, there is another contribution, arising from disturbed fluctuations, which are present even if the mean density remains unchanged. These forces are quantitatively described in a Gaussian field theory~\cite{rohwer2017transient}, and scale for parallel plates with $1/L^d$, as is the case for the equilibrium critical Casimir force. In contrast to the DIFs,  the fluctuation-induced forces relax in time with a power law, due to scale-free relaxation of fluctuation modes parallel to the surfaces. This is a major difference between FIFs and DIFs. We note that, at least for passive particles, the amplitude of the FIF does not depend on the potentials of the immersed surfaces, which is  in contrast to DIF.

The density-induced and fluctuation-induced forces seem largely
decoupled, and the superposition of the two captures the
numerical measurements well, especially for passive Brownian particles, for
which the theory matches the simulation data quantitatively. A quench
in activity for self-propelled particles was shown to lead
qualitatively to the same effects and can thus be seen as quench in
the effective temperature of the active particles, albeit with some
quantitative differences in the amplitudes and time-scales. We
summarize the main properties of DIFs and FIFs in Table
\ref{tab:compare}.

Several ideas exist for future work. Of great interest are more complex time dependencies of $T$, such as periodic variations. Another route is to investigate different surface potentials and the resulting adsorption factors $\alpha$. More generally, the shape dependence of DIFs and FIFs will be interesting to investigate in more detail, also with regard to self-propulsion of non-symmetric objects by DIFs/FIFs.

\subsection{Suggestions for experiments}
There are various possibilities for experimental observation of DIFs and FIFs. Many heating and cooling methods exist for implementing rapid changes of temperature, both in molecular fluids, as well as in suspensions. 

Furthermore, it is interesting to note that, instead of changing $T$, a rapid change in (pair-)potentials is also expected to lead to DIFs and FIFs, in a manner very similar to the phenomena described in this manuscript. Such changes in potentials can be achieved by several methods. For instance, the grafted particles of Ref.~\cite{ballauff2006} drastically change their size by just mild changes of temperature due to a swelling/deswelling transition. The interactions of the paramagnetic particles in the system of Ref.~\cite{maretkeim2004} can be tuned with an external magnetic field, and can thus be switched very quickly over a wide range of strengths.

The advent of active matter in various realizations opens up many more experimental possibilities. The ABPs (``swimmers'') modeled in our simulations have experimental counterparts~\cite{buttinoni2012active,Bechinger16} where, for example, the swimming mechanism of Ref.\cite{buttinoni2012active} is controlled with an external laser field, and can thus be quenched instantaneously. Systems of shaken granular matter~\cite{kudrollikantorkardar2009,kumarramaswamisood2014} are also promising candidates, as the activity may be changed rapidly, e.g. by modifying the shaking protocol. 

\begin{acknowledgments}
  We thank D.~S.~Dean and S.~Dietrich for valuable discussions. This
  work was supported by MIT-Germany Seed Fund Grant
  No. 2746830. C.M.R.~gratefully acknowledges S.~Dietrich for
  financial support. A.S.~acknowledges funding through a PLS
  fellowship of the Gordon and Betty Moore foundation. M.~Kardar is
  supported by the NSF through Grant No.~DMR-1708280. M.~Kr\"uger is
  supported by Deutsche Forschungsgemeinschaft (DFG) Grants No. KR
  3844/2-1 and KR 3844/2-2.

\end{acknowledgments}
\appendix

\section{Adsorption coefficient for plates in a passive ideal gas} 
\label{sec:alpha}
We show here how the coefficient $\alpha$ of Sec.~\ref{sec:twow}, that controls the magnitude of the PQFs induced by density, can be computed for plates embedded in an ideal gas. We consider here the inside of the plates, which are separated by a distance $L$. The plates are modeled by a confining potential $V(z)=V_1(z)+V_2(z)$ where $V_1$ is non-zero only for $z<0$ and $V_2$ only for $z>L$. This is the setup used in simulations with $V_1(z)=\lambda_W z^2/2$ and $V_2(z)=\lambda_W (z-L)^2/2$.

Before the quench, the system is in equilibrium at temperature $T_I$ and bulk density $\rho_0$ so that the density profile reads
\begin{equation}
  \label{eq:app-density-platesI}
  \rho_I(z)=\rho_0 e^{-V(z)/k_B T_I}.
\end{equation}
At infinite time, the system is again in equilibrium at the quench temperature $T_F$ and a different bulk density $\rho_1$ (due to adsorption/desorption), such that
\begin{equation}
  \label{eq:app-density-platesF}
  \rho_F(z)=\rho_1 e^{-V(z)/k_B T_F}.
\end{equation}

Imposing the requirement that particle number is conserved gives the final bulk density as
\begin{equation}
  \label{eq:app-bulk-dens}
  \rho_1=\rho_0\frac{1+2 R_I/L}{1+2 R_F/L}\,,
\end{equation}
where $R_{I/F}=\int_{-\infty}^0 dze^{-V_1(z)/k_BT_{I/F}}$ is a measure
of the characteristic width of the boundary layer near a plate. We
thus get that when $L\gg R_I,R_F$
\begin{equation}
  \label{eq:app-bulk-dens2}
  \rho_1=\rho_0\left[1+\frac{2(R_I-R_F)}{L}+O\left(\frac{R^2}{L^2}\right)\right]\,,
\end{equation}
from which we can read directly the coefficient $\alpha=R_I-R_F$ using Eq.~(\ref{eq:fi}).

\section{Analytical results for the force between two inclusions in a bath of passive, non-interacting particles}
\label{sec:InclAnalyt}

We consider first the one dimensional problem, and later generalize to higher dimensions. In the presence of an inclusion with the potential $V(z)$ at the initial temperature $T_I$, the density is $\rho_I(z) = \rho_0 e^{-\beta_I V(z)}$. After quenching to $T_F$, $\rho(z,t) = \rho_I(z) + \Delta\rho (z,t)$ evolves according to the (density-conserving) Smoluchowski equation~\cite{kreuzer},
\al{
\partial_t \rho(z,t) &= {\Omega} \rho(z,t),
\label{smolua}
}
with ${\Omega} = D_0\left[\partial_z^2 + {\beta_F} \partial_z V'(z) \right]$, subject to $\Delta\rho(z,t=0) =\Delta\rho(z=\pm\infty,t) = 0$, with $D_0 = \mu_0 k_B T$ as before. One finds
\al{
\partial_t \Delta \rho(z,t) &=   
\Omega \Delta \rho(z,t) + \tilde V(z),
\label{rhohat1}
}
where the effective potential $\tilde V(z) = \Omega\rho_I(z)$ must vanish when $T_I = T_F$.

The specific case of a Gaussian inclusion at $z=0$, modeled by the potential in Eq.~\eref{eq:Vgauss1D}, $V(z) = {\frac{V_0 }{\sqrt{2\pi}}}e^{-\frac{z^2}{2\sigma^2}}$ will now be addressed. 
Linearizing Eq.~\eref{rhohat1} in $V_0$ and considering large $T_F$, i.e. $\beta _F \to 0$, one finds
\al{
\partial_t \Delta \rho(z,t) &=  D_0\partial_z^2  \Delta \rho(z,t) \nl
&\qquad + \rho_0{D_0\frac{V_0}{\sqrt{2\pi}}   \beta _I} e^{-\frac{z^2}{2 \sigma ^2}}\frac {(\sigma^2 - z^2)}{\sigma ^4}, 
}
where $\rho_0$ is the homogeneous density in the absence of $V$. 
By using Fourier and Laplace transformations, this equation can be solved analytically, yielding
\al{
\Delta \rho (z,t) = \rho_0\frac{V_0 \beta_I}{\sqrt{2\pi}}\Big[ e^{-z^2/2\sigma^2} - \frac{e^{-\frac{z^2}{2\sigma^2(1+2D_0t/\sigma^2)}}}{\sqrt{1+2D_0t/\sigma^2}}\Big].
\label{rhohat1Dsol}
}
A second inclusion at $z=L$, with the potential $V_{2}(z,L) = {\frac{V_0 }{\sqrt{2\pi}}}e^{-\frac{(z-L)^2}{2\sigma^2}}$, experiences the force 
\al{
F(t) &= \int_{-\infty}^\infty dz \; V_{2}'(z,L) \rho(z,t).
\label{Fgeneral}
}
In $d=1$ this gives
\al{
F(t) &= \rho _0 \frac{\beta _I V_0^2 L}{4 \sqrt{\pi } \sigma }\frac{e^{-\frac{L^2}{4 \sigma^2 \left[ 1 + D_0 t/\sigma^2\right]}}}{\left(1+D_0 t/\sigma^2\right)^{3/2}}.
}
We now identify the change in the size of the inclusion, $\alpha = -\sigma V_0\beta_I$ (using Eq.~\eref{eq:alpha2} to linear order in $V_0$), and the volume of the inclusion, $\mathcal{V}_2=\int dz V_2({z})=\sigma V_0$. Further we express the width of the potentials in terms of the separation of the inclusions, $\sigma \equiv \gamma L$, where $\gamma\ll 1$ is a dimensionless constant. This yields
\al{
F(t) &=  \rho _0 \frac{\alpha \mathcal{V}_2}{ L^2}\frac{e^{-\frac{1}{4 \left(\gamma ^2+ t^*\right)}}}{4 \sqrt{\pi }\left(\gamma ^2+ t ^*\right)^{3/2}},\quad t ^* = {D_0t}/{L^2}.
\label{F1Db}
}

\begin{figure}[t]
\centering
\includegraphics[width=0.99\columnwidth]{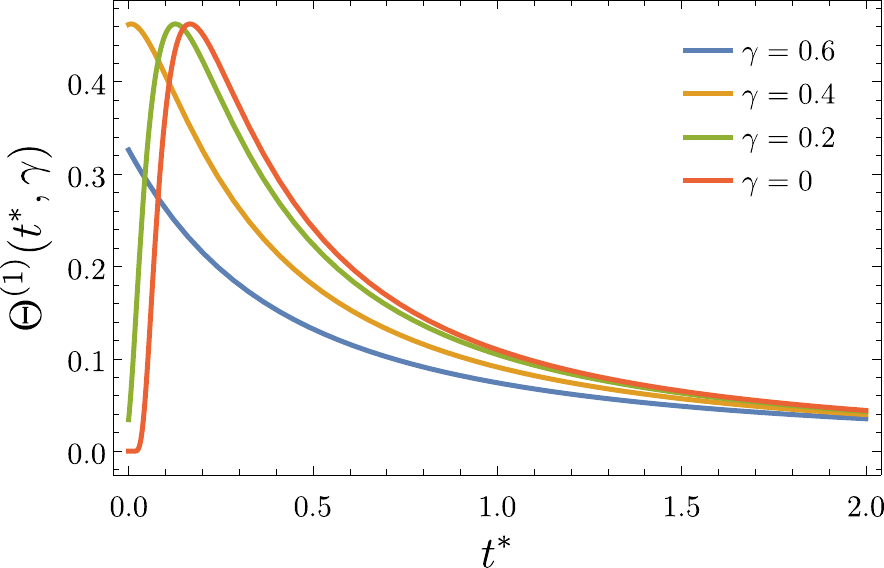}
\caption{The function $\Theta^{(1)}(t^{*},\gamma)$ from \eref{F1Db} (i.e., $d=1$) for various values of $\gamma=\sigma/L$. For $\gamma\neq0$ the force begins at some non-zero value, since the potentials overlap appreciably. As $\gamma\to0$, the curves collapse onto the red master curve, which begins at 0.}
\label{fig:Theta}
\end{figure}

The second term in the 1d solution for $\Delta\rho (x,t)$  in equation \eref{rhohat1Dsol}   
is the solution of the diffusion equation given a Gaussian peak (width = $\sigma$) as initial condition. The extension to $d$ dimensions with radial symmetry ($\bm r \in \mathbb R^d$) is
\al{
\rho_0\frac{V_0 \beta_I}{(2\pi)^{d/2}} \frac{e^{-\frac{r^2}{2\sigma^2(1+2D_0t/\sigma^2)}}}{(1+2D_0t/\sigma^2)^{d/2}},
}
and the corresponding force in $d$ dimensions is
\al{
F(t) &=  \rho _0 \frac{\alpha \mathcal{V}_2}{L^{d+1}} 
\underbrace{
\frac{ e^{-\frac{1}{4 \left(\gamma ^2+ t ^*\right)}}}{2(4\pi)^{d/2}  \left(\gamma ^2+ t ^*\right)^{(d+2)/2}}
}_{\equiv \Theta^{(d)}(t^* ,\gamma)}.
\label{Fd}
}
These analytical results thus confirm and generalize the arguments of Sec.~\ref{sec:two-inclusions}. Indeed, the agreement between Eq.~\eref{eq:Fincl-2} and Eq.~\eref{Fd} is clear for $\gamma=0$. For finite sized inclusions ($\gamma>0$) the time-scale $t^*$ is shifted, as was also observed in simulations (recall \fref{fig:force2d-obstacles}). This is shown for $d=1$ in \fref{fig:Theta}.

\begin{figure}[t!]
  \centering
  \includegraphics[width=0.99\columnwidth]{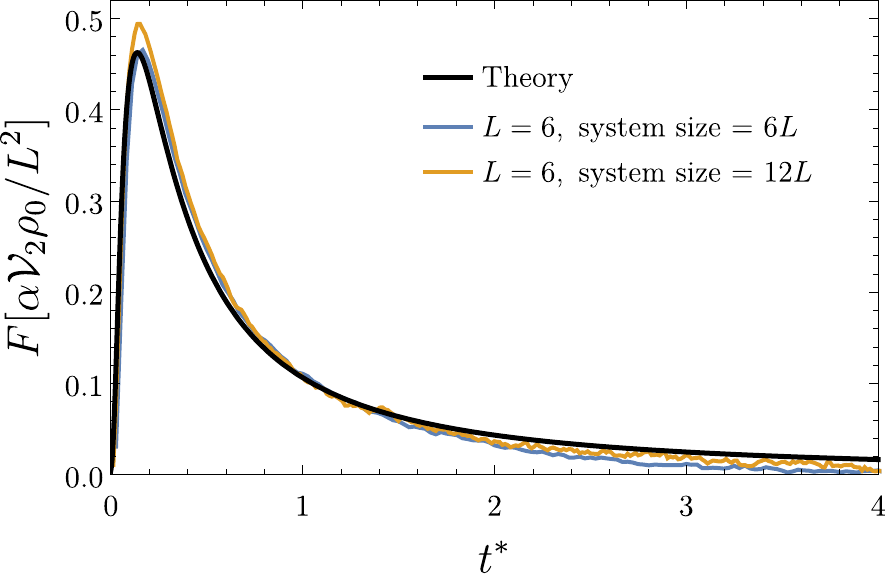}
  \caption{Force between two inclusions modeled by
    Eq.~(\ref{eq:Vgauss1D}) in $d=1$ after a quench from infinite
    temperature to a finite $T_F$ for a passive ideal gas.  The theory
    curve is given by Eq.~(\ref{F1Db}). $\sigma=1$, $L=6$ (i.e., $\gamma=1/6$), $k_BT=5$.}
  \label{fig:force1d-obstacles_convergence}
\end{figure}

\section{Force on inclusions: Convergence with system size}
\label{sec:convergence-inclusions}

We check in Fig.~\ref{fig:force1d-obstacles_convergence} that the
measured force on inclusions converges to the analytical prediction as
the size of the simulation box is increased. At fixed distance $L=6$ between the
inclusions, and high temperatures $k_BT=5$, we see that with
increasing the system size, the measured force approaches the
theoretical prediction. In particular, smaller system sizes act as a
cut-off on the long-time tail of the force.

\clearpage

\newpage

\bibliographystyle{apsrev4-1}
\bibliography{references}

%
%
%
%
%
%
%
%
%
%
%

\end{document}